\documentclass[conference]{IEEEtran}

\usepackage[letterpaper, left=1in, right=1in, bottom=1in, top=0.75in]{geometry}
\usepackage{graphicx}
\usepackage{amsmath}
\usepackage{amsfonts}
\usepackage{subfigure}
\usepackage{multirow}
\usepackage{stfloats}
\usepackage{cite}
\usepackage{color}
\usepackage{graphicx,subfigure,float}
\usepackage{balance}
\usepackage{array}
\usepackage{float}

\usepackage{booktabs}
\usepackage{url}
\usepackage{algorithm}
\usepackage{algorithmic}

\begin{document}

\abovecaptionskip=1pt
\belowcaptionskip=1pt
\belowdisplayskip=2pt
\abovedisplayskip=2pt
\abovedisplayshortskip=0pt
\belowdisplayshortskip=0pt

%
\title{Enabling NB-IoT on Unlicensed Spectrum}

\author{\IEEEauthorblockN{ Rongrong Sun$^1$, Salvatore Talarico$^2$, Wenting Chang$^2$, Huaning Niu$^2$, Hongwen Yang$^1$}
\IEEEauthorblockA{
$^1$Beijing University of Posts and Telecommunications\\
$^2$Intel Mobile Communications Technology Ltd.}
}

\maketitle

\begin{abstract}
 The deployment of Internet of Things (IoT) technologies in unlicensed spectrum is a candidate feature for 5G
to support massive connections of IoT devices. Current IoT unlicensed band technologies, such as Sigfox and LoRa, are all at an early stage of deployment without a significant market share. In this context, the MulteFire (MF) Alliance has envisioned to adapt the cellular NB-IoT design to operate within the Sub-1 GHz unlicensed spectrum. However, the diverse regulatory requirements within this unlicensed band put a hurdle to the world-wide deployment of unlicensed band IoT technologies. To settle this challenge, MF has designed a specific framework for narrow-band (NB)-IoT systems operating on unlicensed spectrum (NB-IoT-U), which can be utilized under both the Federal Communications Commission (FCC) and European Telecommunication Standards Institute (ETSI) regulatory requirements.
 In this paper, enhanced synchronization and physical broadcasting signals are proposed  based upon the framework designed by MF with the aim to allow a more robust detection, and to fulfil the coverage targets set for this technology. Furthermore, in order to allow the system to operate as a frequency hopping spread spectrum (FHSS) system, a novel frequency hopping pattern generator compliant with the regulatory requirements is designed, and its performance is evaluated.
\end{abstract}
\IEEEpeerreviewmaketitle

\section{Introduction}
Recently, demanding of massive applications in various scenarios such as smart cities, smart environments, smart agriculture, and smart hospitals has envisioned the Internet of things (IoT) as a significantly important technology component. In order to enable these scenarios, the Third Generation Partnership Project (3GPP) has been introducing and developing narrowband (NB)-IoT\cite{bio3_NB} technologies starting from Release 13 (Rel-13) with the common objective of enhancing coverage, and battery lifetime while reducing the user equipment (UE) complexity and supporting a massive number of low-throughput devices.

\begin{table*}[t!]
\centering
\caption{Summary of the regulatory requirements for FCC for both FHSS and DTS systems.}
\begin{tabular*}{6.0 in}{|c|c|c|c|c|c|c|}
\hline
Modulation & Frequency Separation & Channel BW &  PSD Limit & EIRP & \# Channels & Dwell Time\\
\hline
\multirow {3}*{FHSS} &\multirow{3}*{ Max\{25kHz,20dB BW\}}&  20dB BW$<250kHz$ & No	&  36	& $\ge$50 & $\le$ 0.4s/20s \\
\cline{3-7}
\multirow {2}*{}&\multirow {2}*{}&\multirow {2}*{$ \text{20dB BW} \in[250, 500]kHz$}& No	& 30& \multirow {1}*{$ \in [25, 50)$} & \multirow {2}*{$\le0.4s/10s$}\\
\cline{4-6}
{}&{}&{}& No &	 36	& $\ge50$&{}\\
\hline
 DTS	& N/A& $\text{6dB BW}\ge500kHz$ &	 8dBm/3kHz & 36&N/A&N/A\\
\hline
\end{tabular*}
\label{Paracontrast}
\vspace{-0.45cm}
\end{table*}

The NB-IoT technology has been designed in 3GPP along the releases to operate in licensed spectrum. However, exponential growth of smart mobile devices connectivity and the deficit of the available spectrum resources in licensed band have resulted in a critical challenge for the capacity of wireless communication systems.  That said, in the past few years a multitude of new radio technologies for IoT, such as LoRa \cite{bio4_Lo}, and Sigfox\cite{bio5_Si}, which benefit from accessing license free spectrum, have emerged. However, in \cite{bio6_MA} the authors evaluated the MAC layer of the LoRa, Sigfox and 3GPP NB-IoT systems, and concluded that the cellular NB-IoT design is more robust than LoRa and Sigfox. Furthermore, in \cite{bio7_Co} the authors simulated the coverage for the same three IoT technologies, and acknowledged the superiority of NB-IoT over LoRa, and Sigfox. Hence, MulteFire (MF) \cite{bio8_MF}, which designed and standardized an LTE-like technology, which solely operates in unlicensed spectrum, is expected to specify the design of an NB-IoT system operating on unlicensed spectrum (NB-IoT-U) with the aim to extend the benefits of the 3GPP NB-IoT design into unlicensed spectrum.

Since path-loss effects are milder at lower frequencies, the unlicensed Sub-1 GHz band results to be very suitable for battery operated IoT devices compared to other unlicensed spectrum, which lie at higher frequencies (i.e., 2.4 GHz and 5 GHz band). However, its Federal Communications Commission (FCC) \cite{bioFCC} and European Telecommunication Standards Institute (ETSI) \cite{bioETSI} regulatory requirements are more complex, and quite different between different geographical regions. To comply with both regulatory bodies, and to cope with the more stringent rules in FCC, MF has designed an hybrid framework allowing to operate the system as both a digital transmission system (DTS) and a frequency hopping spread spectrum (FHSS) system. For this framework, among others it is particularly important to design the synchronization and physical broadcasting signals to fulfil the coverage targets set for this technology. Furthermore, it is important to opportunely design the hopping pattern used when operating as a FHSS system, such that the primary and secondary cell as well as the UEs might be able to know the hopping channel to which the system hops from a set of minimum information. Furthermore, the hopping pattern shall be devised in compliance with the FCC requirements, which impose the pattern to be pseudo random with a uniform distribution across frequency hops.

\section{Contributions and Outline}

In this paper, after providing a thorough overview of the frame structure introduced by MF for NB-IoT-U, we propose a novel algorithm to generate the frequency hopping pattern to use when this operates as a FHSS system, such that this is compliant with the FCC regulatory requirements. Furthermore, we propose a series of enhancement to the discovery reference signal (DRS), including the  physical designs of narrow-band primary and secondary synchronization signals (NPSS/NSSS), and the physical broadcasting channel (NPBCH). The aim of these enhancements is to allow NB-IoT-U to achieve a reliable detection, and fulfill the coverage targets set for this technology, which are in par with the 3GPP NB-IoT design, despite the restrictions (e.g. transmissions power) imposed by the regulatory requirements for the targeted unlicensed band of operation.

While extending the 3GPP NB-IoT design into unlicensed spectrum, since backward compatibility does not need to be supported, there is no need to avoid the usage of the first three OFDM symbols within the subframes (SFs) dedicated for NPSS/NSSS and NPBCH transmission, which in the licensed design are used for physical downlink control channel (PDCCH) transmission. In this matter, new NPSS/NSSS sequences and a new NPBCH resource mapping are designed so that to utilize all the 14 OFDM symbols available within a SF. Furthermore, to further improve performance multiple contiguous time-domain repetitions are applied to these enhanced channels together with a long orthogonal cover code (OCC) used to mitigate false peak detection among NPSS repetitions and channel variations. Additionally, to gain from transmission diversity, a novel multi-antenna precoding vector switching (PVS) transmission scheme is also proposed. As shown along this paper through link-level simulations, all the aforementioned enhancements allow to meet the performance targets set of this technology.


The remainder of this paper is organized as follows. Section \ref{section_2} provides a detailed description of NB-IoT-U frame structure proposed by MF, including a novel method to generate the frequency hopping pattern for such a system. Section \ref{section_3} describes the proposed enhancements to the DRS, which is transmitted on an anchor channel.  In section \ref{section_4}, a performance assessment through link level simulations is  provided for NPSS/NSSS, and NPBCH for the NB-IoT-U framework. This section also provides a thorough evaluation of the frequency hopping generator.  Finally, conclusions are drawn in section \ref{section_5}.

\section{NB-IoT-U System} \label{section_2}

For license free bands, neither the 3GPP NB-IoT half-duplex (HD)-FDD \cite{NB} nor the TDD \cite{bio16_EN} design can be directly reused for the design of a NB-IoT-U due to the limitations imposed by the regulatory bodies.
However, the 3GPP NB-IoT designs should be used as a baseline for the NB-IoT-U design to speed up the standardization process.
While designing NB-IoT-U, these are the key considerations or principles that should be followed:
\begin{itemize}
\item Regulation compliance: The system should faithfully observe the rules imposed by FCC and ETSI on the Sub-1 GHz bands of interest.
\item Regional uniformity: For the purpose of world wide deployment and adoption, and to limit specification impact and complexity of the devices, the frame structure for different regions should be as similar as possible.
\item Development complexity: The downlink (DL) and uplink (UL) numerologies should be inherited from the 3GPP NB-IoT design, so that current cellular chipset implementations can be reused in both the UE, and the eNB.
\item Performance: The system should meet nearly the same performance targets in terms of both maximum coupling loss (MCL), and synchronization latency set for this technology.
\end{itemize}

\begin{figure*}[t!]
\centering
\includegraphics[width=5.8 in, height=1.9 in]{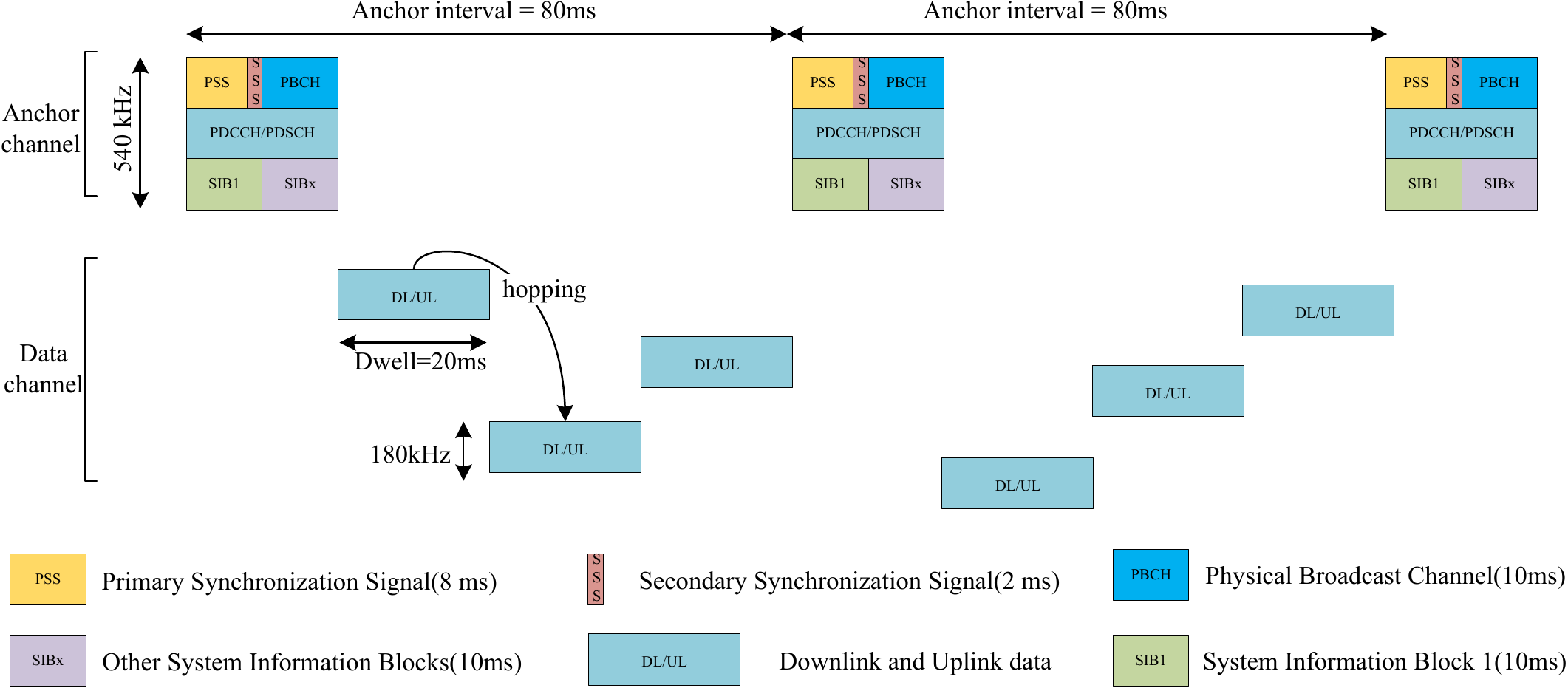}
\caption{NB-IoT-U frame structure to comply with the FCC regulatory requirements.}
\label{framework_fcc}
\vspace{-0.25cm}
\end{figure*}

According to the FCC regulatory requirements \cite{bioFCC}, which are summarized in Table \ref{Paracontrast}, when a system is operated as a DTS system the maximum effective isotropic radiated power (EIRP) is 36 dBm, and the system bandwidth shall be greater than 500 kHz, which provide obvious advantages in terms of capacity, and coverage. On the other hand, a system operated as a FHSS system allows for an higher diversity gain, and more robustness to interference when operating on a smaller bandwidth (i.e, lower than 250 kHz), since the system shall hop on a minimum of 50 channels.
While each operating modality has per se its pros and cons, according to the FCC regulatory requirements a system can be also operated as an hybrid system, meaning that based on the circumstances it can be operated in one mode or in another. In the context of designing a NB-IoT-U system, in order to have a unified bandwidth of 180 kHz, and to allow for frequency diversity and have a more robust design in terms of interference, the UL of an NB-IoT-U system should be operated as a FHSS system. However, on the other hand in order to guarantee higher capacity and coverage the DL of an NB-IoT-U system should be operated as a DTS system. For these reasons, MF has designed the NB-IoT-U system such that it operates as an hybrid system \cite{bio18_AG}.
Owing the simplicity of ETSI regulation requirements, a frame structure compliant with the FCC regulatory requirements can be modified with very minimum changes to become an ETSI-compliant frame structure.



\subsection{FCC-Compliant Frame Structure}

For US and for other countries that need to comply with the FCC regulation, the available band for NB-IoT-U within the Sub-1 GHz band is the 902-928 MHz band. For the FCC-compliant frame structure, the available unlicensed band is partitioned such that to form three anchor channels and 64 data channels. An anchor channel is a dedicated channel over which NPSS/NSSS/NPBCH/ and system information blocks (SIBs) are transmitted, which has a bandwidth of 3 physical resource blocks (PRBs) (i.e., 540 kHz), as illustrated in Fig. \ref{framework_fcc}. This choice is motivated to meet the FCC rules according with 6dB bandwidth should be more than 500 kHz and the power spectral density (PSD) shall be less than 8dBm/3KHz for a system operated as a DTS system. The system is operated as a FHSS system for both DL and UL transmissions with 180 kHz bandwidth. To keep the design of NB-IoT-U in line with the 3GPP NB-IoT design, the DRS, which includes NPSS, NSSS and master information block (MIB) transmission on NPBCH for initial access, is transmitted on the first PRB of each anchor channel. SIBs are carried on the third PRB of each anchor channel, while the second PRB can be occupied by the DL data. To comply with the FCC requirement that states that every channel should be used equally on average, both the anchor channel and all data channel have the same duration of 20 ms. The anchor channel repeats every 80 ms on the same frequency to limit complexity and synchronization latency. The anchor channels appear in fixed positions in both time and frequency domain, while the data channels, which are operated in FHSS mode, need to follow a pseudo-random frequency hopping pattern, as dictated by the regulatory requirements.

\begin{table*}[t]
\centering
\caption{Performance targets of NB-IoT-U and NB-IoT}
\begin{tabular}{c c c c c c c }

\hline \hline
\specialrule{0em}{1pt}{1pt}
  \multicolumn{1}{c}{\textbf{}} &  Spectrum  & Bandwidth (kHz) & Tx Power (dBm)  & Throughput (kbps)&
 MCL (dB) & Battery Life (years) \\
\hline
\specialrule{0em}{1pt}{1pt}
 \multirow {2}*{NB-IoT} & \multirow {2}*{Licensed} & \multirow {2}*{180}	& \multirow {2}*{46/23}	& \multirow {2}*{$<$ 250}	&\multirow {2}*{164} & \multirow {2}*{10}  \\ 
 \specialrule{0em}{2pt}{2pt}
 NB-IoT-U &  \multirow {2}*{Unlicensed} & \multirow {2}*{180}	& \multirow {2}*{36/23}	& \multirow {2}*{$<$ 250}	& \multirow {2}*{161}& \multirow {2}*{10} \\
 (FCC-compliant) \\
 \specialrule{0em}{1pt}{1pt}
  NB-IoT-U & \multirow {2}*{Unlicensed} & \multirow {2}*{180}	& \multirow {2}*{29/23}	& \multirow {2}*{$<$ 250}	& \multirow {2}*{154}& \multirow {2}*{10} \\ 
 (ETSI-compliant) \\
 \specialrule{0em}{1pt}{1pt} \hline
\end{tabular}
\label{Comparison}
 \vspace{-0.35cm}
\end{table*}

\subsection{ETSI-Compliant Frame Structure }

For the EU and all other countries that comply with the ETSI regulation, in an effort for global harmonization within the Sub-1 GHz band, in addition to band 54 (869.4-869.65MHz) a set of bands have been proposed \cite{bio13_ET} to be freed and become soon available for non-specific short-range devices (SRDs), which are suitable for the NB-IoT-U design: for example, band 47b, for which transmissions are only permitted within the bands 865.6-865.8 MHz, 866.2-866.4 MHz, 866.8-867.0 MHz and 867.4-867.6 MHz. For these bands according to the current recommended rules, the maximum allowed EIRP is 29 dBm, and the channel bandwidth must be smaller than 200 kHz. Furthermore, in an effort to mitigate interference, the use of adaptive power control (APC) is required, and the duty cycle is enforced to be smaller than 10\% for network access points, and smaller than 2.5\% for other types of equipment.
Since at the moment the only available band for NB-IoT-U is band 54 with a total of 250 KHz, a single carrier design is developed for NB-IoT-U when compliance with the ETSI regulation is mandated. In an attempt to have regional uniformity, the DRS structure for the ETSI-compliant frame structure is the same as the FCC-compliant frame structure. For this frame structure, in order to comply with the 10\% duty cycle rule, the periodicity of the anchor channel is increased to 1280 ms, and the data for DL and UL is transmitted according to an RRC UL/DL SF configuration, e.g. 8 DL + 72 UL, which can be the same or different for different channels. In future perspective, the system can be operated as a FHSS system, and a frequency hopping patter can be designed as for the design complaint with the FCC regulatory requirements, if more channels will be available in the future.

\subsection{Performance Targets}
Table \ref{Comparison} provides a summary of the performance targets for both 3GPP NB-IoT and NB-IoT-U design. Both technologies target the same battery life. However, the targets MCL are different:
for the FCC-compliant system the target MCL is set by MF \cite{bio18_AG} to 161 dB, which is similar to the 3GPP NB-IoT system, and actually outperforms the competing technologies (e.g. LoRa: 157 dB, Sigfox: 160 dB \cite{coverage}); for the ETSI-compliant system, due to the transmission power limitation (i.e. 29 dBm) imposed by the regulatory requirements, the target MCL is set to 154 dB.
%
%
%

\section{Frequency Hopping Generation}
A Bluetooth system \cite{bio9_BL} operates as a FHSS system, and its hopping sequence is a function of the free-running clock and the first 28 bits of the 48-bit MAC address of the device. By using combinational logic circuits restricted only to XOR, and addition logic functions, the Bluetooth frequency hopping selection kernel is able to generate a unique hopping sequence that is memory-less and pseudo random. In \cite{bio10_PE}, the authors concluded that the Bluetooth hopping sequence offers noticeable residual correlations, and periodic cross correlations features, which lead to uniform usage of the frequency channels over the available band. In light of the advantage of the Bluetooth frequency hopping selection kernel, a similar approach is used to design a novel frequency hopping generator, which is suitable for a NB-IoT-U system, and compliant with the FCC regulatory requirements on this matter.

In order to prevent any memory usage, the frequency hopping patter can be generate based on a permutation operator of a scrambled pre-defined base sequence, similarly as in the Bluetooth design \cite{bio9_BL}. In order to mitigate the inter-cell interferences, it is proposed to uniquely generate the frequency hopping pattern $n_{64}(\text{nSFN})$ as function of the cell ID (PCI) $N_{ID}^{\text{cell}}$, the system frame number (SFN) $\text{nF}$, and the hyper-frame number (HFN) $\text{nHFN}$ using Algorithm \ref{alg1}. In this algorithm, the Perm5(X,P) operator is the same as that defined in the Bluetooth standard, and allows to permutate an input sequence X given a control sequence P. The algorithm is based upon a 32-length sequence $c(i)$, and a 64-length sequence $b(i)$, which are obtained by down-selecting some of the elements from the base hopping sequences for North America defined in the HomeRF \cite{bioHomeRF} and Wi-Fi standard \cite{bioWIfi}, respectively. Both of these two sequences are built to take into account for the additional constraint $\mid b(i)-b(i+1)\mid \ge6$ and $\mid c(i)-c(i+1)\mid \ge6$, inherited for minimizing the co-channel interferences.

\begin{algorithm}[t!]
\footnotesize
\caption{ \small FH pattern generation algorithm}
\label{alg1}
\begin{algorithmic}[1]
\REQUIRE Variables:$N_{ID}^{\text{cell}}$, $\text{nHFN}$, $\text{nF}$.
\ENSURE Frequency Hopping sequence
\STATE  Let $\text{nSFN}$ be defined as follows: \\
 $\text{nSFN}=\lfloor(1024\cdot \text{nHFN}+\text{nF})/2\rfloor $\\
Where $\text{nF}$ is the radio frame number, and $\text{nHFN}$ is the least significant bits of the hyper frame: hyper frame is composed by 1024 radio frames. \\
\STATE Let  $c(i)$, where $ i=1,2,3,\cdots,64$   be a vector defined as follows:\\
 c(i)= \{0,23,62,8,43,16,47,19,61,29,59,22,52,63,26,31,2,18,
 11,36,54,21,3,37,10,34,7,4,60,27,12,25,14,57,41,32,9,58,45,
 20,39,13,33,50,56,42,48,15,5,17,6,49,40,1,28,55,35,53,24,44,
 51,38,30,46\}
\STATE  Let  $b(i)$, where $ i=1,2,3,\cdots,32$   be a vector defined as follows:\\
b(i)=\{0,14,1,16,24,11,22,3,12,13,9,19,5,25,2,17,
      8,23,15,28,10,27,29,21,7,31,6,20,30,4,18,26\}
\STATE Let $n_{32}(\text{nSFN})$be generated as follows:
where  $N(m\hspace{-0.01cm} :\hspace{-0.01cm} n)$ is defined as the vector of bits which represent the  $m^{th}$ bit to the  $n^{th}$ bit in the binary representation of the number $N$
\begin{equation}
\begin{split}
n_{32}(\text{nSFN}) &= \text{Perm5}(X,P)\\
X &=(b(i)+\text{nSFN}(9:5))\text{mod}32\\
i &=\text{nSFN}(4:0)\oplus N_{ID}^{\text{cell}}(4:0)\\
P &=\text{nSFN}(10:5)+64N_{ID}^{\text{cell}}(7:0) \nonumber \\
\end{split}
\end{equation}
\STATE Let $n_{64}(\text{nSFN})$be generated as follows:
\begin{equation}
\begin{split}
n_{64}(\text{nSFN}) &=[c(j)\oplus N_{ID}^{\text{cell}} (5:0)\\
&+\text{nSFN}(10:6)]\text{mod}64\\
j &=[n_{32}(\text{nSFN})+32{\text{nSFN}}_5] \nonumber
\end{split}
\end{equation}
\end{algorithmic}
\end{algorithm}
\setlength{\textfloatsep}{0.15cm}
\setlength{\floatsep}{0.15cm}

\section{Enhanced DRS} \label{section_3}

As shown in Fig. \ref{framework_fcc}, the anchor channel for NB-IoT-U carries the DRS. In order to achieve more robust performances, and allow for lower implementation complexity, while fulfilling the coverage targets set for this technology, in this section a novel DRS structure, and physical design of NPSS/NSSS and NPBCH signals are proposed. In order to optimally utilize the 20 ms length of the anchor channel, the DRS for NB-IoT-U is composed by multiple time-domain repetitions of the NPSS, NPSS and NPBCH signals, which are each one SF long. In order to fulfil the coverage targets for both time/frequency synchronization and MIB detection, it has been identified through an exhaustive simulation campaign that the best choice is to have a DRS composed by eight contiguous NPSS signals, two contiguous NSSS signals, and ten contiguous NPBCH signals.

\subsection{NPSS and NSSS Design}
In NB-IoT-U, multiple synchronization signals are contiguously repeated on a single anchor channel in order to allow coherent combining across time-domain repetitions of the same signal to enhance the detection probability, so initial acquisition can be done with limited latency. In the legacy-LTE design, in order to avoid potential interference between 3GPP NB-IoT and legacy LTE, the first three OFDM symbols of the NB-IoT synchronization signals are punctured by the LTE cell-specific reference signal (CRS). However, for NB-IoT-U backward compatibility with the legacy-LTE is not needed, and the synchronization signals can be enhanced by spanning them over all 14 OFDM symbols within a SF so that to achieve better performance.

\subsubsection{NPSS sequence}
 Similarly as the 3GPP NB-IoT, in NB-IoT-U the proposed NPSS sequence is generated based on a Zadoff-Chu (ZC) sequence, which has inherent constant amplitude zero autocorrelation waveform (CAZAC) properties, and as shown in \cite{robust} provides best estimation for timing offset.
The NPSS sequence for NB-IoT-U is generated in the frequency domain as \cite{TS36}
\begin{equation} \label{eq_1}
d_{l}(n) = S(l)\cdot e^{-j\frac{\pi \mu n(n+1)}{11}}, n=0,1,\cdots,10
\end{equation}

where $n$ is the subcarrier index, $l$ is the index of the OFDM symbol over which the NPSS is mapped into, $\mu$ denotes a root index, and is equal to $\mu=5$ as for the 3GPP NB-IoT design, and $S(l)$ denotes an OCC. While $S(l)$ is a length-11 OCC for the 3GPP NB-IoT design, in NB-IoT-U in order to mitigate false peak detection among the NPSS repetitions within an anchor channel, and channel variations, a long OCC is used, which is given by a binary sequence with length-112 as shown in Table \ref{OCC sequence}. Notice that this OCC has been obtained by simultaneously maximizing the normalized difference between the main and second peak of the autocorrelation of the entire OCC, its first half and its second half, as well as the cross-correlation between the first and second half of the entire OCC. This procedure has been followed in order to design an OCC, which is optimal for both single and multi-antennas transmission when a PVS scheme, as that described later in this paper, is used.

\begin{table}[]
\caption{OCC sequences for NB-IoT-U.}
\renewcommand{\arraystretch}{0.7}
\begin{tabular}{|l|l|}
\hline
\tiny
$S(l)$& \tiny  $l(0),...,l(111)$                                                                                                                                                                                                                                                                                                                       \\ \hline
     & \hfill \begin{tabular}[c]{@{}l@{}}{\tiny[}\tiny -1 1 -1 1 -1 1 -1 -1 1 -1 1 1 1 1 -1 1 -1 -1 -1 -1 1 1 1 -1 1 1 1 -1 1 1 -1 1 1 -1 -1 -1 -1 -1 1 \\ \tiny  -1 -1 1 1 -1 1 1 -1 1 1 1 1 1 -1 1 -1 1 1 -1 -1 -1 -1 -1 -1 1 1 -1 1 1 -1 1 1 1 1 -1 -1 1  -1 -1 -1 \\ \tiny  1 -1 -1 -1 -1 -1 1 1 1 1 -1 -1 -1 -1 1 -1 1 1 1 -1 -1 1 -1 1 1 1 1 -1 1 -1 1 -1 1{]}\end{tabular} \\ \hline
\end{tabular}
\label{OCC sequence}
\end{table}

\subsubsection{NSSS sequence}
For NB-IoT-U, the proposed NSSS sequence is generated from a combination of a frequency domain ZC sequence with 167-length and an Hadamard sequence. This is done in order to span the ZC sequence over exactly 14 OFDM symbols, and allow as in 3GPP NB-IoT design to distinguish the 504 PCIs through the 126 root indices of the ZC sequence, and four orthogonal Hadamard sequences. In particular, the NSSS sequence is generated as
\begin{equation} \label{2}
d(n) = b_{q}(m)\cdot e^{-j2\pi \theta_{x}n}e^{-j\frac{\pi \mu n^{'}(n^{'}+1)}{167}}
\end{equation}
where, $n=0,1,\cdots,167$, $n^{'}= n \text{mod}167$, $m= n \text{mod}160$, and $q = \lfloor N_{ID}^{cell} / 126 \rfloor $. In (2), the term $e^{-j\frac{\pi \mu n^{'}(n^{'}+1)}{167}}$ represents a 167-length ZC sequence, and $\mu$ is a root index that is computed as follow
\begin{equation} \label{3}
\mu=N_{ID}^{cell} \text{mod} 126+3.
\end{equation}
Moreover $e^{-j2\pi \theta_{x}n}$ represents a cyclic shift where the value of $\theta_{x}$ is determined as
\begin{equation} \label{4}
\theta_{x}= \frac{42}{168}(x+1),x=0,1,2,3
\end{equation}
In (\ref{2}), $b_q (m)$ indicates one of the four 160-bits Hadamard sequence, provided in Table \ref{Hadamard sequence}, which are generated so that they are mutually orthogonal among each other to enhance detection probability of the PCI.  Notice that while in 3GPP NB-IoT, the cyclic shift $x$ is used to indicate the 80 ms boundary, for NB-IoT-U design these four cyclic shifts can be used for other purposes, such as paging, and access baring indication.

\subsection{PVS Scheme}
In order to gain from transmission diversity when a multi-antenna transmission is enabled, and be consequently able to further improve the performance for both NPSS and NSSS, a novel PVS based transmission scheme is applied in NB-IoT-U \cite{PVS}. In order to effectively gain from transmission diversity, a precoding vector is changed alternately for every half burst of the NPSS and NSSS repetitions. For instance, for a two antennas transmission, given N the number of NPSS or NSSS consecutive SFs within the DRS, for the first N/2 burst $\mathbf{W}_{2}^{(0)}= {[\begin{matrix} 1 & 1 \end{matrix}]}^{\mathrm{T}}$ is used, while $\mathbf{W}_{2}^{(1)}={[\begin{matrix} 1 & -1 \end{matrix}]}^{\mathrm{T}}$ is used for the remaining SFs.

\begin{table}[]
\caption{Hadamard sequences for NB-IoT-U.}
\renewcommand{\arraystretch}{0.7}
\begin{tabular}{|l|l|}
\hline
\tiny
$q$& \tiny  $b_q(0),...,b_q(159)$                                                                                                                                                                                                                                                                                                                                                                                                                                                                 \\ \hline
\tiny 0 & \begin{tabular}[c]{@{}l@{}}{\tiny[} \tiny1 1 1 1 1 1 1 1 1 1 1 1 1 1 1 1 1 1 1 1 1 1 1 1 1 1 1 1 1 1 1 1 1 1 1 1 1 1  1 1 1 1 1 1 1 1 1  \\  \tiny  1 1 1 1 1 1 1 1 1 1 1 1 1 1 1 1 1 1 1 1 1 1 1 1 1 1 1 1 1 1 1 1 1 1 1 1 1 1 1 1 1 1 1 1 1 1 1 1  \\  \tiny   1 1 1 1  1 1 1 1 1 1 1 1 1 1 1 1 1 1 1 1 1 1 1 1 1 1 1 1 1 1 1 1 1 1 1 1 1 1 1 1 1 1 1 1 1 1 1 1  \\  \tiny  1 1 1 1 1 1 1 1 1 1 1 1 1 1 1 1 1{]}\end{tabular}                                                                                 \\ \hline
\tiny 1 & \begin{tabular}[c]{@{}l@{}}{\tiny[} \tiny 1	-1	-1	1	-1	1	-1	1	1	1	1	-1	-1	1	-1	-1	1	1	-1	-1	-1	1	1	-1	1	-1	1	-1	-1	-1	-1	1	1	-1	1	1	-1	-1	1	1	\\  \tiny  1	-1	-1	1	-1	1	-1	1	1	1	1	-1	-1	1	-1	-1	1	1	-1	-1	-1	1	1	-1	1	-1	1	-1	-1	-1	-1	1	1	-1	1	1	-1	-1	1	1	1	\\  \tiny  -1	-1	1	-1	1	-1	1	1	1	1	-1	-1	1	-1	-1	1	1	-1	-1	-1	1	1	-1	1	-1	1	-1	-1	-1	-1	1	1	-1	1	1	-1	-1	1	1	1	-1	\\  \tiny -1	1	-1	1	-1	1	1	1	1	-1	  -1	1	-1	-1	1	1	-1	-1	-1	1	1	-1	1	-1	1	-1	-1	-1	-1	1	1	-1	1	1	-1	-1	1	1 {]}\end{tabular} \\ \hline
\tiny 2 & \begin{tabular}[c]{@{}l@{}}{\tiny[} \tiny 1	-1	-1	-1	-1	1	-1	1	-1	1	1	1	1	-1	-1	1	-1	-1	1	1	1	-1	-1	-1	-1	1	-1	1	-1	1	1	1	1	-1	-1	1	-1	-1	1	1	\\  \tiny -1	1	1	1	1	-1	1	-1	1	-1	-1	-1	-1	1	1	-1	1	1	-1	-1	-1	1	1	1	1	-1	1	-1	1	-1	-1	-1	-1	1	1	-1	1	1	-1	-1	1	\\  \tiny -1	-1	-1	-1	1	-1	1	-1	1	1	1	1	-1	-1	1	-1	-1	1	1	1	-1	-1	-1	-1	1	-1	1	-1	1	1	1	1	-1	-1	1	-1	-1	1	1	-1	1	\\  \tiny 1	1	1	-1	1	-1	1	-1	-1	-1	-1	1	1	-1	1	1	-1	-1	-1	1	1	1	1	-1	1	-1	1	-1	-1	-1	-1	1	1	-1	1	1	-1	-1{]}\end{tabular} \\ \hline
\tiny 3 & \begin{tabular}[c]{@{}l@{}}{\tiny[} \tiny 1	1	1	-1	-1	-1	-1	1	-1	1	-1	1	1	1	1	-1	-1	1	-1	-1	-1	-1	-1	1	1	1	1	-1	1	-1	1	-1	-1	-1	-1	1	1	-1	1	1	\\  \tiny -1	-1	-1	1	1	1	1	-1	1	-1	1	-1	-1	-1	-1	1	1	-1	1	1	1	1	1	-1	-1	-1	-1	1	-1	1	-1	1	1	1	1	-1	-1	1	-1	-1	1	\\  \tiny 1	1	-1	-1	-1	-1	1	-1	1	-1	1	1	1	1	-1	-1	1	-1	-1	-1	-1	-1	1	1	1	1	-1	1	-1	1	-1	-1	-1	-1	1	1	-1	1	1	-1	\\  \tiny -1	-1	 1	1	1	1	-1	1	-1	1	-1	-1	-1	-1	1	1	-1	1	1	1	1	1	-1	-1	-1	-1	1	-1	1	-1	1	1	1	1	-1	-1	1	-1	-1{]}\end{tabular} \\ \hline
\end{tabular}
\label{Hadamard sequence}
\end{table}

\subsection{NPBCH Design}
In 3GPP NB-IoT \cite{TS36}, the MIB bits are coded into 8 self-decodable code blocks, and each code block is repeated 8 times in 8 consecutive radio frames, which leads to high UE complexity. In NB-IoT-U, the new DRS is structure so that the NPBCH repetitions are consecutive to each other so that coherent combining across repetitions can be applied. In particular, for NB-IoT-U we propose that the MIB bits are coded into 8 self-decodable code blocks, and each code block is repeated in 10 consecutive SFs on the anchor channel. Following the same motivation as for the synchronization signals, PBCH transmission also spans over all the 14 OFDM symbols within a SF.

\section{Performance Evaluation} \label{section_4}

In this section, a performance assessment of the two NB-IoT-U frame structures described in Sec. \ref{section_2} is provided through link level simulations in terms of the detection capabilities for the proposed DRS. Since to the best of the authors knowledge there is absence of relevant research on this topic, and the two frame structures have been only recently developed by MF, it is not yet possible to compare the proposed solutions with other state-of-the-art solutions.  Therefore, the objective of the following section and related results is to demonstrate that the proposed enhancements to the DRS of the NB-IoT-U framework are able to meet the performance requirements set for this technology. Additionally, this section illustrates that the proposed frequency hopping generator for NB-IoT-U is compliant with the regulatory requirements, which mandate that the hopping channels must be nearly uniformly used over time, and for each hopping cycle each channel must be used only once.

%

\begin{table}[t!]
\centering
\caption{Simulation Assumptions.}
\begin{tabular*}{2.4in}{c|c}
\hline
\hline
\textbf{Parameters} & \textbf{Value}\\ \hline
 Channel Model&	TU-1Hz  \\  \hline
 Bandwidth	&180 kHz \\    \hline
 Frequency	Carrier &900MHz   \\    \hline
Antenna Configuration	 &2Tx 1Rx\\ \hline
Residual Frequency Error 	& $\pm 50 Hz$ \\  \hline
Residual Time Error &	 $\pm 64Ts$   \\  \hline
\end{tabular*}
\label{ParaList1}
\end{table}

\subsection{NPSS and NSSS}

\begin{figure}[!t]
\setlength{\abovecaptionskip}{5pt}
\centering
\includegraphics[width=0.45\textwidth]{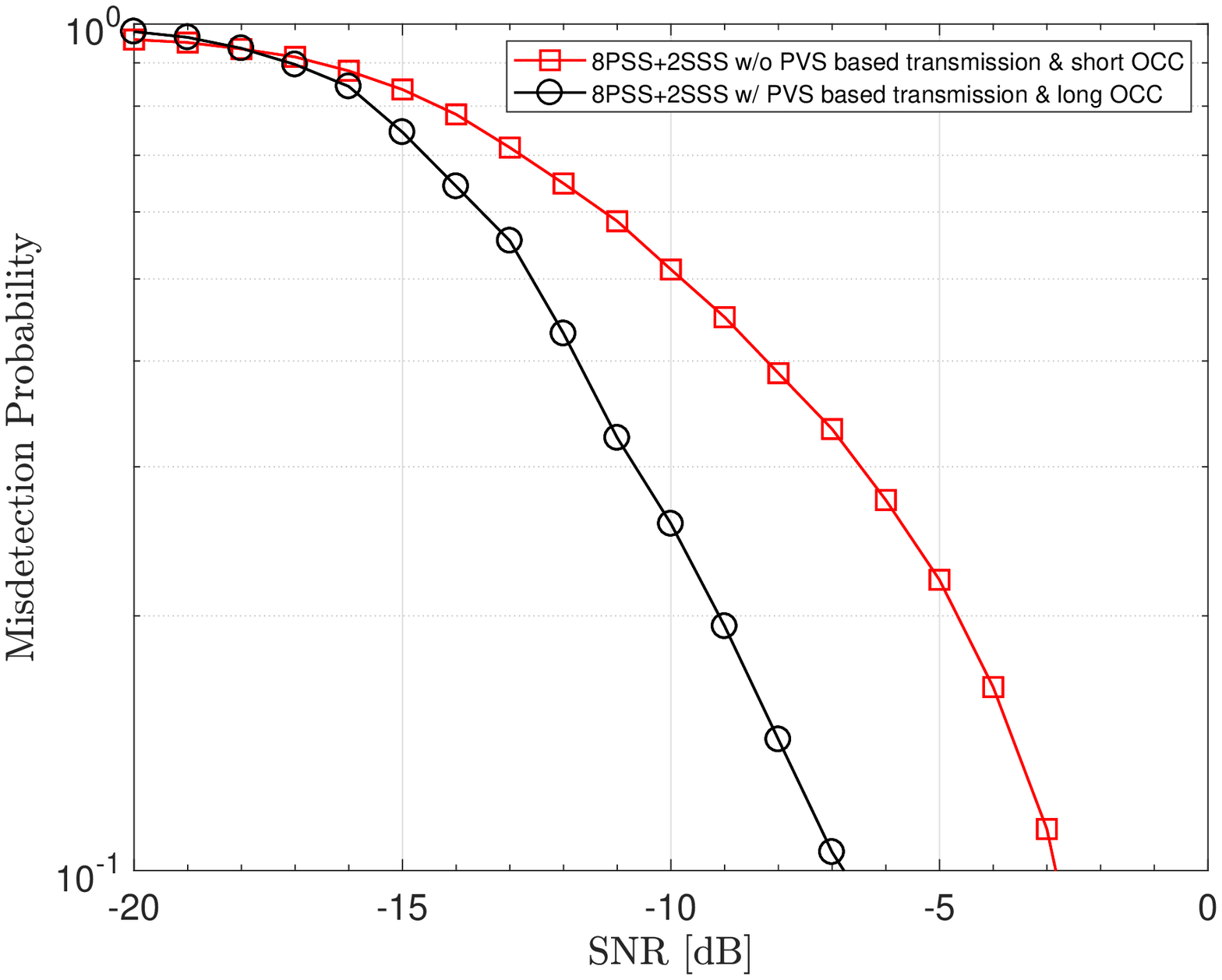}
 \vspace{-0.25cm}
\caption{Synchronization performance over a single anchor channel.}
\label{bler}
\end{figure}

\begin{figure}[!t]
\setlength{\abovecaptionskip}{5pt}
\centering
\includegraphics[width=0.45\textwidth]{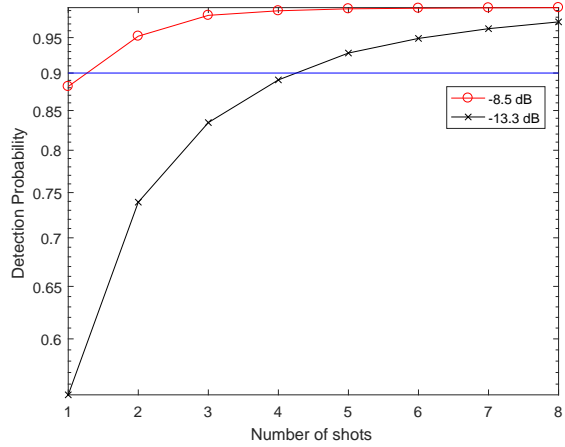}
 \vspace{-0.25cm}
\caption{Synchronization performance when soft-combining is applied across multiple anchor channels or shots.}
\label{detectionprob}
\end{figure}

According to the agreements reached in MF \cite{bio18_AG}, the MCL for NB-IoT operating in Sub-1 GHz band is set to 161 dB (corresponding to a target SNR of -13.3 dB) for countries compliant with the FCC body, and 154 dB (corresponding to a target SNR of -8.5 dB) for European countries, which outperform the competing technologies. By using the simulation assumptions provided in Table \ref{ParaList1}, and by further assuming the initial frequency error is 20 part per million (ppm), the synchronization performance is evaluated. Fig. \ref{bler} shows that a significant improvement can be achieved by using the proposed PVS scheme and long OCC, compared to the case when the proposed PVS scheme is not used and a short OCC is applied to the NPSS repetitions as in 3GPP NB-IoT design. Furthermore, this figure shows that  90\% detection confidence can be achieved by a required SNR of nearly -7 dB through the detection of a single anchor channel. This emphasizes that in order to meet the requirements that MF has set soft-combining across multiple anchor channels is needed. However, Fig. \ref {detectionprob} shows that soft-combining across only five anchor channels is needed to reach 90\% detection confidence for countries compliant with the FCC regulation body, and only 2 attempts can reach 90\% detection probability for countries compliant with the ETSI regulation body. Based on Fig. \ref {detectionprob}, it is possible to infer that
 for the FCC-compliant frame structure the synchronization latency is approximately 400 ms, which is even better than the 3GPP NB-IoT typical synchronization latency (520 ms for stand-alone deployment \cite{NB}). However, for ETSI-complaint frame structure, due to the long periodicity between two anchors, which is dictated by the 10\% duty cycle imposed by the regulatory requirements, the synchronization latency is 2560 ms.

\begin{figure}[!t]
\setlength{\abovecaptionskip}{5pt}
\centering
\includegraphics[width=0.45\textwidth]{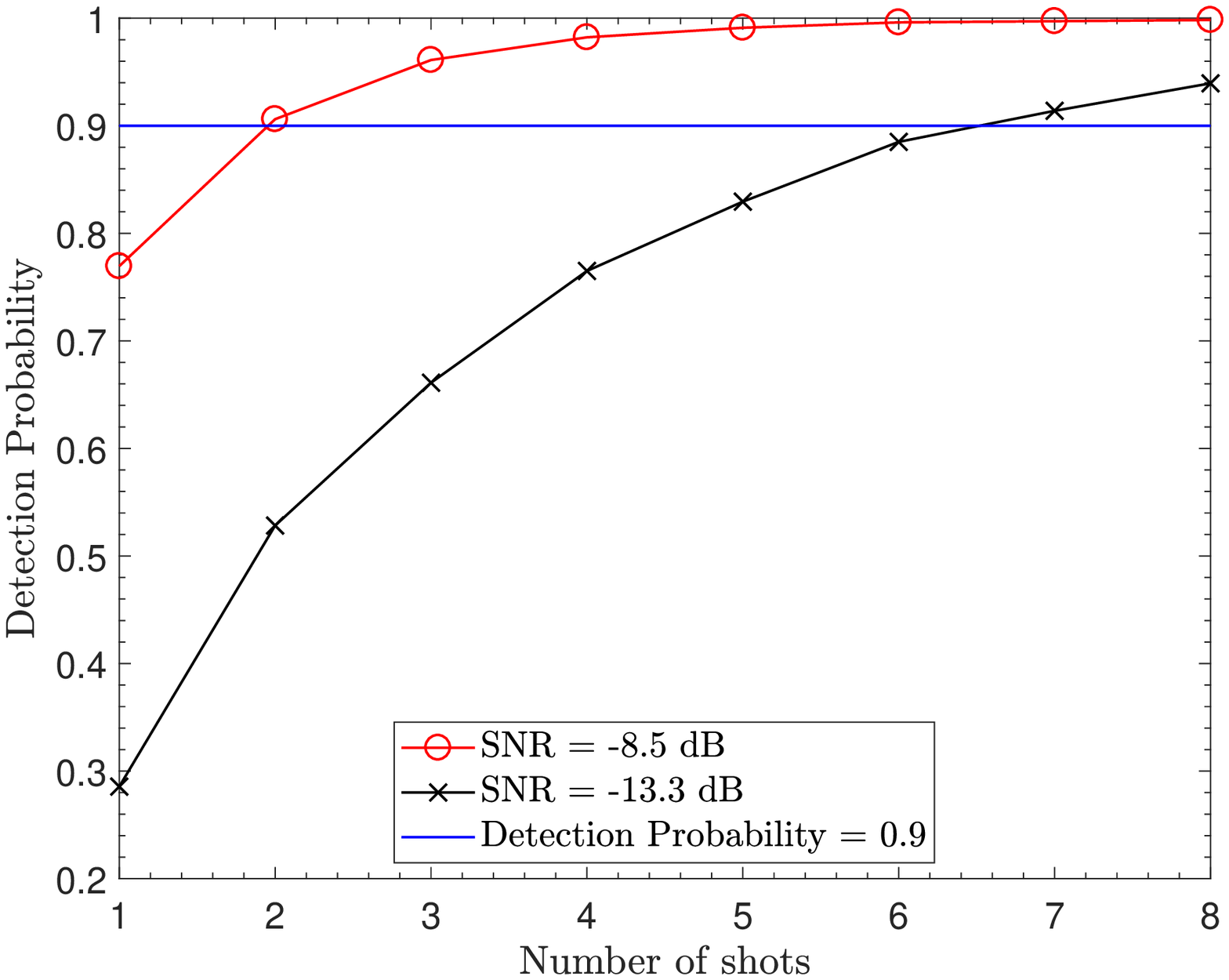}
 \vspace{-0.15cm}
\caption{PBCH performance when soft-combining is applied across multiple anchor channels or shots.}
\label{PBCH2}
\end{figure}

\subsection{PBCH}
 By using the simulation assumption summarized in Table \ref{ParaList1}, the performance of the proposed NPBCH for NB-IoT-U is evaluated. Fig. \ref{PBCH2} shows the detection probability when soft-combining across multiple anchor channels is used as function of the number of anchor channel or shots for the target SNR for countries complaint to the FCC and ETSI regulation. This figure highlights that in order to meet the requirements that MF has set soft-combining across multiple shots is needed. However, only two shots are needed for countries compliant with the ETSI regulation body, and 7 shots are needed for countries compliant with the FCC regulation body, which in this case corresponds to 560 ms delay for MIB acquisition that is less than the typical 640 ms for the 3GPP NB-IoT design \cite{NB}.

\subsection{Frequency Hopping Pattern}

\begin{figure}[!t]
\setlength{\abovecaptionskip}{5pt}
\centering
\includegraphics[width=0.45\textwidth]{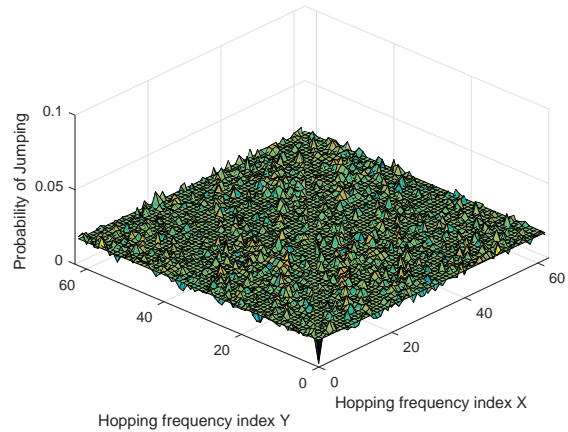}
\caption{Probability of jumping across frequency channels.}
\label{jumpP}
\end{figure}

In order to evaluate the performance of the proposed frequency hopping generator, Fig. \ref{jumpP} shows the probability that by jumping from one hop to the next hop we land in the hopping frequency index X given that we were in the hopping frequency index Y. In the ideal case of uniform distributed channels, the probability should be always equal to 1/(N-1), where  N  is the number of channels. As shown by Fig. \ref{jumpP} the jumping probability for all combinations of channels is close to the ideal case of 1/63, given that N = 64, and is zero for the case when $X=Y$, meaning that jumping on the same channel does not occur.

\section{Conclusion}  \label{section_5}

In this paper, a new framework with compliance to both the FCC and the ETSI regulatory requirements for cellular NB-IoT systems operating on the Sub-1 GHz unlicensed band is described. An enhanced DRS design for this type of framework is proposed. Link level simulations are provided to demonstrate that the proposed enhancements the NB-IoT-U framework is able to meet the performance requirements set for this technology, which outperforms competing technologies. In this paper, a new frequency hopping generator is also proposed to accommodate the system to operate as a FHSS system. By evaluating the statistics of the proposed frequency hopping generator, it has been proved that this is compliant with the regulatory requirements, and the channels are nearly uniformly used.

\balance

\end{document}